\documentclass[twocolumn,aps,pra,showpacs,floatfix]{revtex4}

\usepackage{dcolumn}
\usepackage{graphicx}
\usepackage{epsf}
\usepackage{amsmath}
\usepackage{bm}
\usepackage{times}

\def\half{{\textstyle{1\over 2}}}

\def\half{{\textstyle{\frac12}}}
\def\LZa{{\ln\left[\half(Z\alpha)^{-2}\right]}}
\def\LZasquared{{\ln^2\left[\half(Z\alpha)^{-2}\right]}}

\newcommand{\addrHD}{
Max--Planck--Institut f\"ur Kernphysik,
Saupfercheckweg 1, 69117 Heidelberg, Germany}

\newcommand{\addrNIST}{
National Institute of Standards and Technology,
Gaithersburg, Maryland 20899--8401, USA}

\begin{document}

\sloppy

\title{Two--Loop Bethe Logarithms for non-$\boldsymbol S$ Levels}

\author{Ulrich D.~Jentschura}
\affiliation{\addrHD}
\affiliation{\addrNIST}

\begin{abstract}
Two-loop Bethe logarithms are calculated 
for excited $P$ and $D$ states in hydrogenlike systems, and estimates
are presented for all states with higher angular momenta. 
These results complete our knowledge of the $P$ and $D$ energy levels 
in hydrogen at the order of $\alpha^8\,m_{\rm e}\,c^2$,
where $m_{\rm e}$ is the electron mass and $c$ is the 
speed of light, and 
scale as $Z^6$, where $Z$ is the 
nuclear charge number. Our analytic and numerical calculations
are consistent with the 
complete absence of logarithmic terms 
of order $(\alpha/\pi)^2\,(Z\alpha)^6\,
\ln[(Z\alpha)^{-2}]\,m_{\rm e}\,c^2$ for $D$ states and 
all states with higher angular momenta.
For higher excited $P$ and 
$D$ states, a number of poles from lower-lying levels
have to subtracted in the numerical evaluation.
We find that, surprisingly,
the corrections of the ``squared decay-rate type'' are the numerically
dominant contributions in the order 
$(\alpha/\pi)^2\,(Z\alpha)^6\,m_{\rm e}\,c^2$ for states with large
angular momenta, and provide an estimate of the
entire $B_{60}$-coefficient for Rydberg states
with high angular momentum quantum numbers. Our results
reach the predictive limits of the quantum electrodynamic
theory of the Lamb shift.
\end{abstract}


\maketitle

%
%
\section{INTRODUCTION}
\label{intro}

The two-loop Bethe logarithm stems from a nonrelativistic 
treatment of the full two-loop self-energy. It is the finite
part which is left over when the two-loop problem is 
renormalized according to the original derivation of 
Bethe~\cite{Be1947}
for the one-loop self-energy and therefore represents
a natural generalization of the most basic quantum electrodynamic 
calculation to the two-loop level, yet the 
respective calculation is the only way to complete the analysis
of hydrogenic energy levels in the order 
$(\alpha/\pi)^2 \, (Z\alpha)^6 m_{\rm e}\,c^2$. 
In comparison to the one-loop problem, the two-loop nonrelativistic
self-energy is much more complicated, and it involves three instead
of one propagator denominators. Matrix elements cannot be expressed in 
closed analytic form. Numerical methods, inspired by lattice
calculations, represent convenient tools for the evaluations.

Final calculations were carried out on workstation clusters 
of the Max--Planck--Institute for Nuclear Physics
in Heidelberg and of the National Institute
of Standards and Technology. The total CPU time of all computations
required for the present paper was about 30 months,
but the real time was of course drastically
shortened due to parallel processing. The two-loop problem is 
known to be computationally demanding.
Additional difficulties arise for excited 
states due to a number of poles from lower-lying levels
which have to be subtracted in the numerical evaluation.
After subtraction, the result is finite, as it should be, and
represents an observable energy shift.

The problem of so-called squared decay rates is associated with the 
two-loop Bethe logarithm~\cite{JeEvKePa2002}. The
squared decay rates, which originate from specific
photon energies where two propagators become singular
simultaneously, cannot be 
uniquely interpreted as energy corrections even if the 
concept of a pole on the second sheet of the Riemann surface 
defining the propagator is used in order to define energy levels,
and have been assigned~\cite{JeEvKePa2002}
to an contribution $\delta B_{60}$,
where $B_{60}$ is the two-loop coefficient multiplying the 
scaling factor of $(\alpha/\pi)^2 \, (Z\alpha)^6 m_{\rm e} c^2/n^3$
(here, $n$ is the principal quantum number). 
For $3P$ and $4P$ states, the squared decay rates were originally
evaluated in in Ref.~\cite{JeEvKePa2002}. In~\cite{Je2004b60}, 
an analysis of $S$ states was supplemented, and it has
been clarified that the contributions as given 
in~\cite{JeEvKePa2002} should be understood in 
radians per second rather than 
cycles per second, or Hz. 
Here, we supply numerical values for the correction of 
the ``squared-decay'' type for excited $S$, $P$, $D$, 
$F$, $G$ and $H$ states, and we also discuss approximations
for general states. 
The squared decay rates are a manifestation of the 
fact that a proper definition of energy levels ceases to be possible
at the level of $\alpha^2 (Z\alpha)^6$ in units of the 
electron mass, as already pointed out in 
Ref.~\cite{Lo1952}. Yet, these contributions are
mathematically well defined and have to be evaluated 
in a calculation whose aim is to explore the predictive 
limits of the quantum electrodynamic theory of 
hydrogenlike systems.

This paper is organized as follows. In Sec.~\ref{known}, we
summarize all known results for the two-loop energy shift
of excited $P$ and $D$ states up to the order $\alpha\,(Z\alpha)^6$
(all energy shifts are measured in units of $m_{\rm e}\,c^2$ in this 
paper unless stated otherwise). Important definitions and formulas
regarding the two-loop Bethe logarithm are recalled in 
Sec.~\ref{twlp}. The asymptotics of the two-loop integrand and
the calculation of the two-loop Bethe logarithm are 
discussed in Sec.~\ref{asympcalc}. The squared decay rates, which are
numerically significant especially for states with higher 
angular momenta, are treated in Sec.~\ref{square}.
Miscellaneous two-loop results are compiled in 
Appendixes~\ref{miscP}---\ref{miscFGH}. One-photon vacuum-polarization 
and self-energy shifts 
of order $\alpha (Z\alpha)^7$ are discussed in Appendix~\ref{vp}.

%
%
\section{Known Two--Loop Results}
\label{known}

We work in natural units ($\hbar = c = \epsilon_0 = 1$),
as is customary in QED bound-state calculations,
and we also set the electron mass equal to unity in the 
following.
The real part of the energy displacement of a hydrogenic state
due to the two-loop (2L) self-energy (SE) can be parameterized as follows,
\begin{equation}
\label{DefESE2L}
\Delta E^{\rm (2L)}_{\mathrm{SE}} = \left(\frac{\alpha}{\pi}\right)^2 \,
\frac{(Z\alpha)^4}{n^3} \, H(Z\alpha)\,,
\end{equation}
where $H$ is a dimensionless function.
Here, we are concerned with $P$ and $D$ states, and in part with 
states of higher angular momenta. For these manifolds, 
the first nonvanishing terms in the semi-analytic expansion
of the dimensionless function
$H(Z\alpha)$ in powers for $Z\alpha$ and $\ln(Z\alpha)$ 
read as follows,
\begin{eqnarray} 
\label{defH}
H(Z\alpha) &=& B_{40} 
+ (Z\alpha)^2 \biggl\{ B_{62} \ln^2[(Z\alpha)^{-2}]
\nonumber\\[2ex]
&& + B_{61} \ln[(Z\alpha)^{-2}] + B_{60} \biggr\}.
\end{eqnarray}
The first index of the $B$-coefficients denotes the power of
$Z\alpha$ including the factors $Z\alpha$ contained in 
Eq.~(\ref{DefESE2L}), 
and the second index denotes the power of $\ln[(Z\alpha)^{-2}]$. 
For the lowest-order $B_{40}$ term, a complete
result is known which is valid for any hydrogenic 
state (see, e.g., Ref.~\cite{MoTa2005}).
We here express the result in terms
of the expectation value of 
an operator, as in Eq.~(8.1) of Ref.~\cite{JeCzPa2005},
and suppress all operators which are zero for $P$ states
and states with higher angular momenta,
\begin{eqnarray}
\frac{(Z\alpha)^4}{n^3} \, B_{40} &=& 
\left[ \frac{197}{288}
- \frac{3}{2}\,\zeta(2)\,\ln 2
+ \frac{1}{4}\,\zeta(2)
+ \frac{3}{8}\,\zeta(3)
\right] 
\nonumber\\[2ex]
&& \times \left< \sigma^{ij}\,\nabla^i V\,p^j \right>\,.
\nonumber
\end{eqnarray}
We adopt here all conventions of Ref.~\cite{JeCzPa2005},
in particular $\sigma^{ij} = [\sigma^i,\,\sigma^j]/(2\,{\rm i})$
for the $(2 \times 2)$-dimensional spin matrices.
We recall the well-known identity
\begin{equation}
\left< \sigma^{ij} \, \nabla^i V \, p^j \right> =
\frac{(Z\alpha)^4}{n^3}\, 
\left( - \frac{2}{\kappa \, (2l +1)} \right)\,,
\end{equation}
where 
$\kappa = (-1)^{j+l+1/2} \, \left(j + \frac12 \right)$ is the 
Dirac angular quantum number ($l$ is the orbital angular
momentum quantum number, and $j$ denotes the total
angular momentum of the electron).
In this way, one immediately reproduces the well-known general result
\begin{eqnarray}
\label{b40nonS}
B_{40} &=& - \frac{2}{\kappa \, (2l +1)}\,
\nonumber\\[2ex]
& & \times \left[ \frac{197}{288}
- \frac{3}{2}\,\zeta(2)\,\ln 2
+ \frac{1}{4}\,\zeta(2)
+ \frac{3}{8}\,\zeta(3)
\right]\,,
\end{eqnarray}
which is valid for all non-$S$ hydrogenic states.

Our goal here is to extend formula (\ref{b40nonS}) to
two relative orders of $Z\alpha$, for 
general $P$ and $D$ states, and to find estimates 
for general states with higher angular momenta.
In Ref.~\cite{JeCzPa2005}, numerical results have
only been indicated for the fine-structure difference 
of $D$ states [see Eqs.~(4.22) and (5.3) of Ref.~\cite{JeCzPa2005}] 
and for $P$ states [see Eqs.~(4.24) and (5.4), (6.2) and 
(7.7) of Ref.~\cite{JeCzPa2005}]. However, the complete 
result for $B_{60}$ entails the two-loop Bethe logarithm,
which has not been known so far.

We recall that 
in terms of matrix elements, to be evaluated on Pauli--Schr\"{o}dinger
nonrelativistic wave functions of non-$S$ hydrogenic states,
the higher-order terms $B_{62}$, $B_{61}$ and $B_{60}$ can be 
expressed as follows (see Ref.~\cite{JeCzPa2005}),
\begin{widetext}
\begin{align}
\label{b626160}
\lefteqn{\frac{(Z\alpha)^6}{n^3} 
\left\{ B_{62}\ln^2[(Z \alpha)^{-2}] + B_{61}\ln[(Z \alpha)^{-2}] + B_{60} 
\right\} 
= \frac{(Z\alpha)^6}{n^3} \,
\left\{ b_L + \beta_{4} + \beta_{5} + 
\left[\frac{38}{45} + 
\frac{4}{3} \LZa{} \right]\, N \right\} }
\nonumber \\ 
& + \left[-\frac{42923}{259200} 
+ \frac{9}{16} \zeta(2) \ln (2)
- \frac{5 \zeta(2)}{36} 
- \frac{9 \zeta(3)}{64} 
+ \frac{19}{135} \LZa
+ \frac{1}{9} \LZasquared \right]
\left< \vec{\nabla}^2 V \frac{1}{(E-H)'} \vec{\nabla}^2 V \right> 
\nonumber \\ 
& + \left[ \frac{2179}{10368} 
- \frac{9}{16} \,\zeta(2)\,\ln (2)
+ \frac{5}{36} \,\zeta(2)
+ \frac{9}{64}\,\zeta(3) \right]\,
\left< \vec{\nabla}^2 V\,\frac{1}{(E-H)'}  \, \vec{p}^{\,4} \right> 
\nonumber \\ 
& + \left[ -\frac{197}{1152} 
+ \frac{3}{8} \,\zeta(2)\,\ln (2)
- \frac{1}{16} \,\zeta(2)
- \frac{3}{32}\,\zeta(3) \right]\,
\left< \vec{p}^{\,4} \,\frac{1}{(E-H)'} \, \sigma^{ij}\,\nabla^i V\,p^j \right> 
\nonumber \\ 
& + \left[ \frac{233}{576} 
- \frac{3}{4} \,\zeta(2)\,\ln (2)
+ \frac{1}{8} \,\zeta(2)
+ \frac{3}{16}\,\zeta(3) \right]\,
\left< \sigma^{ij}\,\nabla^i V\,p^j \,\frac{1}{(E-H)'} \, 
\sigma^{ij}\,\nabla^i V\,p^j \right> 
\nonumber \\ 
& + 
\left[-\frac{197}{2304} 
+ \frac{3}{16} \,\zeta(2)\,\ln (2)
- \frac{1}{32} \,\zeta(2)
- \frac{3}{64}\,\zeta(3)\right]\,
\left< 
\left\{ \vec{p}^{\,2} , \vec{\nabla}^2 V + 2\,\sigma^{ij}\,
\nabla^i V\,p^j\right\}
\right> 
\nonumber \\ 
& + \left[-\frac{83}{1152}  
+ \frac{17}{8} \zeta(2)\,{\ln(2)}
- \frac{59}{72}\,\zeta(2) 
- \frac{17}{32}\,\zeta(3) \right]\,
\left< \left(\vec{\nabla} V \right)^2\right>
\nonumber \\ 
& + \left[-\frac{87697}{345600} 
+ \frac{9}{10}\zeta(2)\,\ln (2)
- \frac{2167}{9600}\zeta(2)
- \frac{9}{40}\zeta(3)
+ \frac{19}{270} \LZa
+ \frac{1}{18} \LZasquared \right]\,
\left< \vec{\nabla}^4 V \right>
\nonumber \\ 
& + \left[-\frac{16841}{207360}  
- \frac{1}{5}\, \zeta(2)\,\ln(2)
+ \frac{223}{2880} \, \zeta(2)
+ \frac{1}{20}\,\zeta(3)
+ \frac{1}{24}\, \LZa
\right]\,
\left< 2\,{\rm i}\,\sigma^{ij}\, p^i\,\vec{\nabla}^2V\,p^j\right>\,.
\end{align}
\end{widetext}
We take the opportunity to point out that in the 
corresponding Eq.~(8.1) of Ref.~\cite{JeCzPa2005},
a prefactor of $(\alpha/\pi)^2$ in front of the 
second term on the right-hand side was missing
[in Ref.~\cite{JeCzPa2005}, the entire above result
(\ref{b626160}) for the energy shift
was multiplied by a factor of $(\alpha/\pi)^2$ on the left
and right-hand side].
Also, the square of the logarithm $\LZasquared$ 
in the second and in the last-but-one term 
on the right-hand side were not included.
Both typographical errors are absent from 
Eq.~(8) of previous work reported in 
Ref.~\cite{CzJePa2005} and from the above Eq.~(\ref{b626160}).

We recall here the definitions of the quantities $N$,
$\beta_{4}$ and $\beta_{5}$, which enter into Eq.~(\ref{b626160}).
In doing so, we first recall that in accordance with 
the notation introduced in~\cite{CzJePa2005},
we redefine the finite part of an integral which diverges as 
\begin{subequations}
\label{defFP}
\begin{equation}
\label{defFP1}
\int\limits_0^\Lambda dk\,f(k) = A \, \lambda + 
B \, \ln\lambda + C + {\cal O}(\lambda^{-1})\,, 
\end{equation}
with $\lambda\equiv \Lambda\,(Z\,\alpha)^{-2}$, 
for a specified upper cutoff $\Lambda$, to be
equal to just the constant term, i.e. 
\begin{equation}
\label{defFP2}
\int_0^\Lambda dk\,f(k) \equiv C.
\end{equation}
\end{subequations}
With this definition in mind, we have
\begin{eqnarray}
\frac{(Z\alpha)^6 \, N}{n^3} &=& 
\frac{2}{3} Z\,\alpha\, \int_0^{\Lambda}\: dk k\,
\delta_{\pi\,\delta^3(r)}
\left< \,\vec p\,\frac{1}{E-H-k}\,\vec p\,\right>, 
\nonumber\\[0.5ex]
\frac{(Z\alpha)^6 \,\beta_4}{n^3} &=& 
\frac{2}{3} \int_0^{\Lambda} \: dk k
\delta_{(\sigma^{ij}\nabla^i\,V\,p^j/4)}
\left< \vec p\frac{1}{E-H-k}\vec p\right>, 
\nonumber\\[0.5ex]
\label{defbetaN}
\frac{(Z\alpha)^6 \,\beta_{5}}{n^3} &=& 
-\frac{1}{2} \int_0^\Lambda \: dk k\,
\left< \sigma^{ij}\nabla^j V\frac{1}{E-H-k} p^i \right>.
\end{eqnarray}
The $N$ term has previously been defined in
Refs.~\cite{Pa2001,PaJe2003}; it is generated by a Dirac delta correction
to the Bethe logarithm. The notation $\delta_V$ is used 
in accordance with Ref.~\cite{Pa2001,JeCzPa2005} in order
to denote the first-order perturbation of 
the Hamiltonian, the energy and the wave function in the 
ensuing matrix element, due to the specified potential $V$.

The evaluation of the two-loop 
Bethe logarithm for $1S$ and $2S$ has been discussed in
Ref.~\cite{PaJe2003}, and for $3S$--$6S$ in Ref.~\cite{Je2004b60}.
For $1S$ and $2S$, there is no ambiguity in the definition
of the Bethe logarithm, because the integration over both 
photon energies in the nonrelativistic self-energy is 
free of singularities.
However, for all higher excited $S$ states and 
all states considered here, one incurs 
real (rather than imaginary) contributions to the energy shift
from the product of imaginary contributions due to singularities 
along both photon integrations, and these result in 
``squared decay rates'' in the sense of Ref.~\cite{JeEvKePa2002}.
Thus, it is helpful to make a clear distinction between the 
singularity-free, principal-value part $\overline{b}_L$ and 
a real part $\delta^2 B_{60}$, which is incurred by 
``squared'' (or, more precisely, products of) 
imaginary contributions from the pole terms. We write
\begin{equation}
\label{bLdef}
b_L = \overline{b}_L + \delta^2 B_{60}\,,
\end{equation}
where $\overline{b}_L$ is obtained as the nonlogarithmic 
energy shift stemming from the nonrelativistic 
self-energy, with all integrations carried out by principal 
value, and $\delta^2 B_{60}$ is the corresponding 
contribution defined in Refs.~\cite{JeEvKePa2002,Je2004b60}, 
due to squared imaginary parts.
The exact meaning of the separation (\ref{bLdef}) will be
clarified below.
Here, we just note that for $3S$--$6S$ states, the above 
separation is not really essential, because $\delta^2 B_{60}$
is a numerically marginal contribution as compared to 
$\overline{b}_L$ (see Ref.~\cite{Je2004b60}), 
and thus $b_L(nS) \approx \overline{b}_L(nS)$ to a 
very good approximation.
For the states under investigation here, the distinction (\ref{bLdef}),
surprisingly, proves to be highly essential; yet before we come 
to a discussion of this surprising phenomenon, let us first 
discuss the evaluation of $\overline{b}_L$.

%
%
\section{Two--Loop Bethe Logarithm}
\label{twlp}

We briefly recall~\cite{PaJe2003,Je2004b60} the nonrelativistic 
two-photon self-energy $\Delta E_{\rm NRQED}$ 
as an integral of the following structure,
\begin{equation}
\label{NRQED}
\Delta E_{\rm NRQED} = 
\left( \frac{2 \alpha}{3 \pi m^2} \right)^2
\int_0^{\epsilon_1} dk_1 \, k_1 
\int_0^{\epsilon_2} dk_2 \, k_2 \, f(k_1, k_2)\,,
\end{equation}
where the $k_1$ and $k_2$ represent photon energies,
and $\epsilon_1$ as well as $\epsilon_2$ are cutoff 
parameters. The function $f(k_1, k_2)$ is defined
in Eq.~(\ref{fk1k2}) below; its precise structure is 
unimportant for the following consideration,
which concerns the relation of the 
cutoff parameters $\epsilon_i$ ($i=1,2$) to the 
ultraviolet cutoff $\Lambda$ parameter used in 
Eqs.~(\ref{defFP})---(\ref{defbetaN}).

In order to clarify this relation,
we recall that in the context of the $\epsilon$ method (see \S 123 of
Ref.~\cite{BeLiPi1982} and~\cite{Pa1993,JePa1996,JePa2002}), 
the cutoff parameters are chosen so that the 
$\epsilon$ can be made 
arbitrarily small, but only under the condition
$Z\alpha \ll \epsilon$, so that the expansion,
first carried out in $Z\alpha$, then in $\epsilon$,
for the low- as well as the high-energy parts, 
gives the complete result for the self-energy.
This procedure has been fully clarified in
\S 123 of Ref.~\cite{BeLiPi1982} and in
Refs.~\cite{Pa1993,JePa1996,JePa2002,Je2005mpla}.
In Ref.~\cite{Je2005mpla}, it has been stressed that this method actually
corresponds to an expansion in {\em large} $\epsilon$,
and indeed, in the context of the dimensional regularization
method~\cite{JeCzPa2005}, the nonlogarithmic term which 
remains after subtraction of the divergent contributions 
as $\epsilon_1 = \Lambda_1 \to \infty$, 
$\epsilon_2 = \Lambda_2 \to \infty$, has
been been identified as the two-loop Bethe logarithm $b_L$.
Because we are dealing here with excited states that can decay via dipole
radiation, care must be taken in the definition of the 
integration prescription for the $k_1$ and $k_2$ integrations.
In the current section, we first assume a principal-value
prescription and write 
\begin{align}
\label{bLint}
& \frac{(Z\alpha)^6}{n^3} \, \overline{b}_L = \\
& \qquad \frac{4}{9} ({\rm P.V.}) \int_0^{\Lambda_1} {\rm d}k_1 k_1 
({\rm P.V.}) \int_0^{\Lambda_2} {\rm d}k_2  k_2 f(k_1, k_2),
\nonumber
\end{align}
where $({\rm P.V.})$ denotes the principal value, 
and it is understood that divergent terms for large $\Lambda_1$
and $\Lambda_2$ have to be subtracted, in the sense of 
Eq.~(\ref{defFP}).

The subtractions for large $\Lambda_i$ ($i=1,2$) 
are carried out assuming 
$\Lambda_1 \ll \Lambda_2$, so we first let $\Lambda_2 \to \infty$,
extract the constant term as a function of $k_1$
and then we integrate this term with respect to $k_1$,
letting $\Lambda_1 \to \infty$.
In this way we ``sweep'' the entire first quadrant of the 
two-dimensional $(k_1, k_2)$ plane (i.e., the entire region
$k_1 > 0$, $k_2 > 0$).
Of course, the same result would be obtained for the constant term
under the opposite sequence of first letting 
$\Lambda_1$ approach infinity, then $\Lambda_2$.

We are now in the position
to recall the explicit form of $f(k_1, k_2)$, which reads
\begin{widetext}
\begin{align}
\label{fk1k2}
& f(k_1, k_2) =
-\left< p^i \, \frac{1}{H - E + k_1} \, p^j \, 
\frac{1}{H - E + k_1 + k_2} \, p^i \, 
\frac{1}{H - E + k_2} \, p^j \right> 
\nonumber\\[1ex]
& - \frac{1}{2} 
\left< p^i \frac{1}{H - E + k_1}  p^j 
\frac{1}{H - E + k_1 + k_2} p^j 
\frac{1}{H - E + k_1} p^i \right> 
- \frac{1}{2} 
\left< p^i \frac{1}{H - E + k_2} p^j 
\frac{1}{H - E + k_1 + k_2} p^j 
\frac{1}{H - E + k_2} p^i \right>
\nonumber\\[1ex]  
& - 
\left< p^i \frac{1}{H - E + k_1} p^i 
\left( \frac{1}{H - E} \right)' p^j
\frac{1}{H - E + k_2} p^j \right> 
+ \frac{1}{2} \,
\left< p^i \, \frac{1}{H - E + k_1} \, p^i \right> \,
\left< p^j \, \left( \frac{1}{H - E + k_2} \right)^2 \, p^j \right>
\nonumber\\[1ex]  
& + \frac{1}{2} \,
\left< p^i \, \frac{1}{H - E + k_2} \, p^i \right> \,
\left< p^j \, \left( \frac{1}{H - E + k_1} \right)^2 \, p^j \right>
+ \left< p^i \, \frac{1}{H - E + k_1} \, 
\frac{1}{H - E + k_2} \, p^i \right>
\nonumber\\[1ex]  
& + \frac{1}{k_1 + k_2} \,
\left< p^i \, \frac{1}{H - E + k_2} \, p^i \right>
+ \frac{1}{k_1 + k_2} \,
\left< p^i \, \frac{1}{H - E + k_1} \, p^i \right> \,.
\end{align} 
\end{widetext}
The interpretation of the terms on the right-hand side
is as follows: the first six are due to fourth-order
perturbation theory generated by the ``velocity-gauge''
nonrelativistic ($\vec{p}\cdot \vec{A}$)-interaction.
Herein, the fifth and the sixth terms are derivative terms which 
naturally occur in fourth-order perturbation theory.
The seventh term involves (on the left and on the right)
($\vec{p}\cdot \vec{A}$)-interactions, and a seagull term 
proportional to $\vec{A}^2$ in between.
The eighth and ninth terms involve a seagull term 
outside of the ($\vec{p}\cdot \vec{A}$)-interactions.
We also recall that the scaling of 
$f(k_1, k_2)$ with $Z\alpha$ is completely clarified in
Ref.~\cite{Je2004b60}.

%
%
\section{ASYMPTOTICS OF THE INTEGRAND}
\label{asympcalc}

We first consider the asymptotics relevant for the initial
$k_2$ integration in Eq.~(\ref{bLint}).
In the limit $k_2 \gg k_1$, we find
\begin{align}
& k_2\,f(k_1, k_2) = 
\left< p^i \, \frac{H-E}{(H - E - k_1)^2} \, p^i \right>
\\[2ex]
& +  \frac{1}{k_2} \, 
(Z\,\alpha) \, \delta_{\pi \delta^3(r)}
\left< p^i \, \frac{1}{E - H - k_1} \, p^i \right>
+ {\cal O}\left(\frac{1}{k_2^2}\right)\,.
\nonumber
\end{align}
The second term vanishes for $D$ states and all
states with higher angular momenta.
Subtracting the two above terms according to the 
definition (\ref{defFP}) [a practical procedure is
outlined in Eq.~(26) of Ref.~\cite{Je2004b60}],
we obtain 
\begin{equation}
g(k_1) = ({\rm P.V.}) \int_0^{\Lambda_2} dk_2 \,
k_2 \, f(k_1, k_2) \,.
\end{equation}
We now have to calculate
\begin{equation}
\frac{(Z\alpha)^6}{n^3} \, \overline{b}_L = 
\frac{4}{9} ({\rm P.V.}) \int_0^{\Lambda_1} dk_1 \, k_1 \, g(k_1) \,,
\end{equation}
and this necessitates the calculation of the asymptotics 
of $g(k_1)$ for large $k_1$. After a rather long 
and tedious calculation, we find
\begin{equation}
\label{k1asymp}
k_1 \, g(k_1) = - 2 \, \ln k_0 + \frac{3 N}{2 k_1} + 
{\cal O}\left(\frac{1}{k_1^2}\right)\,,
\end{equation}
a result which is valid only for non-$S$ states.
The $N$ term, which is defined in Eq.~(\ref{defbetaN}),
remains important for the $P$ state calculation, but 
vanishes for $D$ states and all states of higher angular momenta.

In the actual calculations,
we use a lattice formulation of the Schr\"{o}dinger propagator 
\cite{SaOe1989} with up to $200\,000$ lattice sites.
This provides sufficient accuracy (and computational efficiency)
for all calculations reported here and is feasible on 
average-size workstation clusters. 

It is computationally advantageous to isolate the 
contribution to $\overline{b}_L$ due to the fourth 
term on the right-hand side of Eq.~(\ref{bLdef}), in
order to avoid numerical problems associated with 
the calculation of the reduced Green function. 
This contribution reads, after the $k_1$ and $k_2$-integrations
and appropriate subtractions,
\begin{align}
\label{loop-after-loop}
& \frac{(Z\alpha)^6}{n^3}\, L_2 = 
\frac49 \, \sum_{\phi_m \neq \phi} \frac{1}{E - E_m}
\nonumber\\[2ex]
& \times \left| \left< \phi \left| 
p^i (H-E) \ln\left[\frac{|H-E|}{(Z\alpha)^2}\right] p^i 
\right| \phi_m \right> \right|^2\,,
\end{align}
where $\phi$ is the reference state.
The calculation of this contribution is done separately,
by writing it in terms of a sum over virtual states, which 
are calculated as a basis set on a lattice in coordinate space.
Numerical values are compiled in Table~\ref{table1},
where the name ``loop-after-loop'' should be taken 
{\em cum grano salis} because the negative-energy virtual 
states are excluded from expression (\ref{loop-after-loop}).
The final results for $\overline{b}_L$ are in Table~\ref{table2}.

Based on the trend of the data in Table~\ref{table2},
we conjecture here that 
in the limit $n \to \infty$, the two-loop Bethe logarithms 
for a given $l$ should approach a constant in the same way as the 
one-loop Bethe logarithms~\cite{Po1981,DrSw1990,JeMo2005bethe}.
Also, the data in Table~\ref{table2} indicate that the two-loop 
Bethe logarithms, just like their one-loop counterparts, become 
smaller in magnitude for increasing orbital angular momentum.
Indeed, they do so quite drastically, with the $D$-state values
being almost two orders of magnitude smaller than the 
$P$-state logarithms. The magnitude of the $S$-state 
logarithms~\cite{Je2004b60}, in turn,
is in the range of $60 \dots 80$ and thus larger than 
the $P$-state values by more than one order of magnitude.

%
%
\section{SQUARED DECAY RATES}
\label{square}

As clarified in Ref.~\cite{JeEvKePa2002} and Sec.~IV
of Ref.~\cite{Je2004b60}, squared decay rates represent natural
limits to which energy levels can be uniquely associated 
with a particular atomic level. Corrections of this type 
have been analyzed for $2P$ and $3P$ in Ref.~\cite{JeEvKePa2002},
and for $3S$ and $4S$ in Ref.~\cite{Je2004b60}.
Here, we supplement values for $5P$ and $6P$, as well as 
all other states with $n \leq 6$.
Some inaccuracies in previous treatments  for 
$S$ and $P$ states
are corrected in Table~\ref{table3}, and a more extensive 
list of levels is covered.

\begin{widetext}
\begin{center}
\begin{table}[htb]
\begin{center}
\begin{minipage}{13.6cm}
\caption{\label{table1}
Loop-after-loop contribution $L_2$ to the 
two-loop Bethe logarithm for $P$, $D$ and $F$ states.
For the $6H$ state, the result
for the loop-after-loop contribution
reads $-0.000(1)$, implying a negative sign.}
\begin{tabular}{c@{\hspace{0.2cm}}c@{\hspace{0.5cm}}%
c@{\hspace{0.2cm}}c@{\hspace{0.5cm}}c@{\hspace{0.2cm}}%
c@{\hspace{0.5cm}}c@{\hspace{0.5cm}}c@{\hspace{0.5cm}}c@{\hspace{0.5cm}}c}
\hline
\hline 
\rule[-2mm]{0mm}{6mm}
level & $L_2$ & level & $L_2$ & level & $L_2$ & level & $L_2$ & level & $L_2$ \\
\hline
$2P$ & $-0.97(2)$ & $-$  & $-$         & $-$  & $-$         & $-$  & $-$         & $-$  & $-$ \\ 
$3P$ & $-1.15(2)$ & $3D$ & $-0.004(1)$ & $-$  & $-$         & $-$  & $-$         & $-$  & $-$ \\ 
$4P$ & $-1.19(2)$ & $4D$ & $-0.004(1)$ & $4F$ & $-0.002(1)$ & $-$  & $-$         & $-$  & $-$ \\  
$5P$ & $-1.23(2)$ & $5D$ & $-0.005(1)$ & $5F$ & $-0.002(1)$ & $5G$ & $-0.001(1)$ & $-$  & $-$ \\  
$6P$ & $-1.26(3)$ & $6D$ & $-0.005(1)$ & $6F$ & $-0.002(1)$ & $6G$ & $-0.001(1)$ & $6H$ & $-0.000(1)$\\
\hline
\hline
\end{tabular}
\end{minipage}
\end{center}
\end{table}
\end{center}
\end{widetext}

\begin{figure}[htb]
\begin{center}
\begin{minipage}{8.6cm}
\begin{center}
\includegraphics[width=1.0\linewidth]{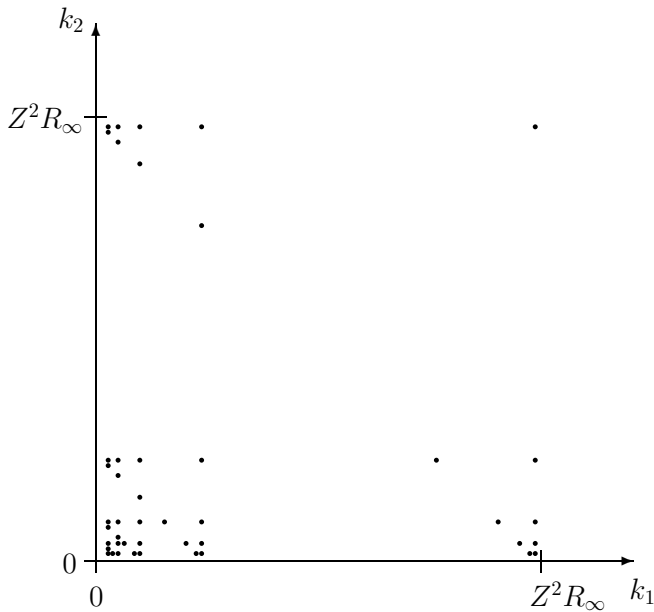}
\caption{\label{fig1} Location of the 
discrete pairs of frequencies $(k_1, k_2)$ which give rise
to the squared decay rates for an $n=6$ state.
The point $Z^2 R_\infty = (Z\alpha)^2 m/2$ denotes a ``$Z$-dependent
Rydberg constant'' and marks the ionization
energy of a nonrelativistic hydrogenlike atom system.}
\end{center}
\end{minipage}
\end{center}
\end{figure}

According to Refs.~\cite{JeEvKePa2002,Je2004b60},
the term $\delta^2 B_{60}$ is generated by ``squared decay rates,''
which correspond to well-defined, isolated points of
the $(k_1, k_2)$-photon energy plane, where
two propagators become singular simultaneously in
the integrand of the two-loop Bethe logarithm given in
Eq.~(\ref{fk1k2}). These are all points where, for a given 
principal quantum number $n$ of the reference state,
any two of the following conditions are satisfied
simultaneously,
\begin{align}
k_1 =& -(Z\alpha)^2 \, \left( \frac{1}{2n^2} - \frac{1}{2m_1^2} \right)\,,
\\[2ex]
k_2 =& -(Z\alpha)^2 \, \left( \frac{1}{2n^2} - \frac{1}{2m_2^2} \right)\,,
\\[2ex]
k_1 + k_2 =& -(Z\alpha)^2 \, \left( \frac{1}{2n^2}-\frac{1}{2m^2} \right)
\,,
\end{align}
where $m_1, m_2, m < n$ are values of the principal quantum
numbers in the intermediate states. In Fig.~\ref{fig1}, we list 
all of these points in the two-dimensional $(k_1, k_2)$-plane 
for a state with a principal quantum number $n = 6$.
All frequencies which give rise to the singularities are
smaller than $(Z \alpha)^2/2 = Z^2 R_\infty$.

Upon picking up the squared imaginary 
contribution, we obtain for the following structures
from the two-loop integrand~(\ref{fk1k2}), 
\begin{subequations}
\label{prescription}
\begin{align}
\label{polea}
\int_0^{\Lambda_1} dk_1 \int_0^{\Lambda_2} dk_2
\frac{1}{k_1 - a} \, \frac{1}{(k_2 - b)^2} \to 0\,,
\\[2ex]
\label{poleb}
\int_0^{\Lambda_1} dk_1 \int_0^{\Lambda_2} dk_2
\frac{1}{(k_1 - a)^2} \, \frac{1}{k_2 - b} \to 0\,,
\\[2ex]
\label{polec}
\int_0^{\Lambda_1} dk_1 \int_0^{\Lambda_2} dk_2 
\frac{1}{k_1 - a} \, \frac{1}{k_2 - b} \to - \pi^2\,,
\\[2ex]
\label{poled}
\int_0^{\Lambda_1} dk_1 \int_0^{\Lambda_2} dk_2
\frac{1}{k_1 - a} \, \frac{1}{k_1 + k_2 - b} \to - \pi^2\,.
\end{align}
\end{subequations}
One might ask why the result on the right-hand side reads
$-\pi^2$, not $-(2\pi)^2$. The answer is that 
both the $k_1$ and the $k_2$ integration contours
have to be deformed along {\em half} circles,
centered at the location of the singularities, infinitesimally
below the real axis, i.e.~in the mathematically
positive sense. This half circle entails a factor $\pi {\rm i}$, whose
square gives the results on the right-hand sides 
of Eqs.~(\ref{polea})---(\ref{poled}).

Finally, the total squared decay rate
\begin{equation}
\delta^2\Gamma = \left( \frac{\alpha}{\pi} \right)^2 \,
\frac{(Z\alpha)^2}{n^3} \, \delta^2 B_{60}
\end{equation}
is obtained 
after picking up all terms which result as the product of two 
imaginary parts, when integrating over both $k_1$ and $k_2$ in 
Eq.~(\ref{fk1k2}), according to the prescription 
(\ref{prescription}). The contributions $\delta^2 B_{60}$ is
thus mathematically well-defined 
and, in particular, has an unambiguously defined sign for 
each individual state.

A very curious observation can be made regarding the 
order-of-magnitude of $b_L$ versus that of $\delta^2 B_{60}$
for different manifolds of hydrogenic states.
For $S$ and $P$ states, we have $|\delta^2 B_{60}| \ll |b_L|$,
as is evident from Table~\ref{table2} of this work and from 
Table II of Ref.~\cite{Je2004b60}.
For the $1S$ and the $2S$ states considered in 
Ref.~\cite{PaJe2003},
we even have $\delta^2 B_{60}(1S) = \delta^2 B_{60}(2S) = 0$.
For higher excited $S$ states, the magnitude 
of $\delta^2 B_{60}$ is smaller than the magnitude of 
the current numerical uncertainty of $\overline{b}_L$ 
incurred by the numerical 
integration described in Ref.~\cite{Je2004b60},
and thus, $\delta^2 B_{60}$ can at present be completely
ignored for $S$ states.
Therefore, the question of whether to include 
$\delta^2 B_{60}$ into the canonical definition of 
$b_L$ had been ignored in Refs.~\cite{PaJe2003,Je2004b60} 
for the simple reasons
that $\delta^2 B_{60}(nS)$ was either vanishing or
found to be a numerically negligible.
Initially (see Ref.~\cite{JeEvKePa2002}),
the notation $\delta^2 B_{60}$ was chosen such as to
make an association with a small ``uncertainty''
in the definition of $b_L$.
However, the situation changes drastically for $P$
states, where $|\delta^2 B_{60}| \approx 0.1 \, |b_L|$
for $n \geq 3$, and even more drastically for $D$ states,
where $|\delta^2 B_{60}| \gg |b_L|$ for all $n$, reversing the 
hierarchy of the ``uncertainty'' $\delta^2 B_{60}$
and the two-loop Bethe logarithm $b_L$
(the latter would otherwise
be assumed to constitute the dominant effect).

In view of this situation, the pressing question arises 
whether or not one should include $\delta^2 B_{60}(nS)$ into the 
definition of the two-loop Bethe logarithm for states with 
higher angular momenta. One might argue that 
$\delta^2 B_{60}(nS)$ should be excluded from the 
definition of the two-loop Bethe logarithm 
$b_L$ because, as shown in Ref.~\cite{JeEvKePa2002}, the term cannot be 
uniquely interpreted as an energy shift associated with 
a specific atomic level. On the other hand, 
one might argue that
$\delta^2 B_{60}(nS)$ should be included into the definition of the 
two-loop Bethe logarithm $b_L$ for higher
excited states, because it is generated
by the product of two imaginary energy shifts, both of 
which originate from the same nonrelativistic 
two-loop self-energy (\ref{NRQED}), which otherwise gives rise to
the low-energy contribution to the $B_{60}$
coefficient. We reemphasize 
that the $\delta^2 B_{60}$-term has an unambiguously 
defined sign for each individual state 
and is important for the comparison of our analytic
approach to $B_{60}$ to any nonperturbative
(in $Z\alpha$) numerical calculation
of the two-loop energy shift for the listed excited
hydrogenic states, as well as for the comparison of theory
and experiment based on a line-shape formalism~\cite{Lo1952}. 
Nonperturbative calculations of the two-loop
self-energy have recently been pursued
intensively~\cite{MaSa1998a,MaSa1998b,YeSh2001,YeInSh2003,%
YeInSh2005jetp,YeInSh2005,YeInSh2006}.

In the current investigation, we would like to 
follow the second route and define [see Eq.~(\ref{bLdef})]
$b_L = \overline{b}_L + \delta^2 B_{60}$,
where $\overline{b}_L$ is obtained by performing
all integrations in Eq.~(\ref{bLint}) via a principal-value prescription,
with the results listed in Table~\ref{table2}, and $\delta^2 B_{60}$
contains all the terms generated by the products of 
imaginary parts incurred at poles in the $(k_1, k_2)$-plane.
This definition has the following advantages:
{\em (i)} When the definition (\ref{bLdef}) is adopted,
the term $b_L$ in the result for the $B_{60}$ coefficient as
given in Eq.~(8.1) of Ref.~\cite{JeCzPa2005} contains
all real (rather than imaginary) terms which 
can be inferred from the nonrelativistic self-energy given 
in Eq.~(16) of Ref.~\cite{Pa2001} and Eq.~(5) of 
Ref.~\cite{PaJe2003} upon contour integration
over both photon energies. 
{\em (ii)}  In nonperturbative numerical evaluations of radiative 
two-loop energy shifts for excited states, one would
``pick up'' terms of the type $\delta^2 B_{60}$ naturally
if one deforms both photon energy integration contours 
according to the Feynman prescription near adjacent 
bound-state poles. The definition (\ref{bLdef}) makes possible
a direct comparison of nonperturbative numerical results obtained in 
this way, to the results for $b_L$ and $B_{60}$ given here,
without the need for any subtraction of the 
squared imaginary parts.
{\em (iii)} For $D$ states, the definition (\ref{bLdef}) includes the 
numerically most significant real (rather than imaginary) 
contribution which can be inferred from the nonrelativistic 
two-loop self-energy (\ref{NRQED}) into $b_L$ and thus into 
the $B_{60}$ coefficient.

Despite these advantages of the proposed definition (\ref{bLdef}),
one should remember that the term $\delta^2 B_{60}$ cannot be 
interpreted as an energy shift uniquely associated with a
particular atomic level, and has to be treated differently,
namely according to ideas outlined in Sec.~3 of 
Ref.~\cite{JeEvKePa2002} when in a comparison to experiments.
According to the discussion in the cited reference, 
a valid way of treating (part of) $\delta^2 B_{60}$
would first entail a subtraction of this term from from the energy shift,
and then, a reinterpretation of it in terms of an off-diagonal
``decay rate operator'' which has to be included into a 
line-shape formalism.

\begin{widetext}
\begin{center}
\begin{table}
\begin{center}
\begin{minipage}{17.0cm}
\caption{\label{table2}
Numerical values for the principal-value integrated contribution
$\overline{b}_L$ to the two-loop Bethe logarithms for $P$, $D$ and $F$ states.
Values for $S$ states are taken from~\cite{Je2004b60} 
and are listed for completeness. For the $6H$ state, the result
for $\overline{b}_L$ reads $-0.000(1)$, implying a negative sign.}
\begin{tabular}{c@{\hspace{0.3cm}}c@{\hspace{0.7cm}}%
c@{\hspace{0.3cm}}c@{\hspace{0.7cm}}c@{\hspace{0.3cm}}c@{\hspace{0.7cm}}c@{\hspace{0.3cm}}c@{\hspace{0.7cm}}%
c@{\hspace{0.3cm}}c@{\hspace{0.7cm}}c@{\hspace{0.3cm}}c}
\hline
\hline 
\rule[-2mm]{0mm}{6mm}
level & $\overline{b}_L$ & level & $\overline{b}_L$ & level & $\overline{b}_L$ & level & $\overline{b}_L$ &
level & $\overline{b}_L$ & level & $\overline{b}_L$ \\
\hline
$1S$ & $-81.4(3)$ & $-$  & $-$       & $-$  & $-$          & $-$  & $-$         & $-$  & $-$         & $-$  & $-$        \\
$2S$ & $-66.6(3)$ & $2P$ & $-2.2(3)$ & $-$  & $-$          & $-$  & $-$         & $-$  & $-$         & $-$  & $-$        \\ 
$3S$ & $-63.5(6)$ & $3P$ & $-2.5(3)$ & $3D$ & $-0.006(2)$  & $-$  & $-$         & $-$  & $-$         & $-$  & $-$        \\ 
$4S$ & $-61.8(8)$ & $4P$ & $-2.8(3)$ & $4D$ & $-0.004(2)$  & $4F$ & $-0.002(1)$ & $-$  & $-$         & $-$  & $-$        \\  
$5S$ & $-60.6(8)$ & $5P$ & $-2.8(3)$ & $5D$ & $-0.005(3)$  & $5F$ & $-0.002(1)$ & $5G$ & $-0.001(1)$ & $-$  & $-$        \\  
$6S$ & $-59.8(8)$ & $6P$ & $-2.9(3)$ & $6D$ & $-0.006(4)$  & $5F$ & $-0.002(1)$ & $6G$ & $-0.001(1)$ & $6H$ & $-0.000(1)$ \\
\hline
\hline
\end{tabular}
\end{minipage}
\end{center}
\end{table}
\end{center}

\begin{center}
\begin{table}
\begin{center}
\begin{minipage}{16.0cm}
\caption{\label{table3}
Numerical values for the correction $\delta^2 B_{60}$
for $S$, $P$, $D$ and $F$ states. Values for $3S$ and $4S$ have
already been discussed in Ref.~\cite{Je2004b60}.
As explained in the text, the 
contributions listed below cannot be unambiguously interpreted
as energy shifts associated with a particular 
atomic level, although they are mathematically
well defined.}
\begin{tabular}{c@{\hspace{0.3cm}}c@{\hspace{0.7cm}}%
c@{\hspace{0.3cm}}c@{\hspace{0.7cm}}%
c@{\hspace{0.3cm}}c@{\hspace{0.7cm}}%
c@{\hspace{0.3cm}}c@{\hspace{0.7cm}}%
c@{\hspace{0.3cm}}c@{\hspace{0.7cm}}%
c@{\hspace{0.3cm}}c}
\hline
\hline
\rule[-2mm]{0mm}{6mm}
level & $\delta^2 B_{60}$ & level & $\delta^2 B_{60}$ &
level & $\delta^2 B_{60}$ & level & $\delta^2 B_{60}$ &
level & $\delta^2 B_{60}$ & level & $\delta^2 B_{60}$ \\
\hline
$1S$ & $0.0$  & $-$  & $-$      & $-$ & $-$          & $-$ & $-$       & $-$ & $-$       & $-$ & $-$ \\
$2S$ & $0.0$  & $2P$ & $-0.008$ & $-$ & $-$          & $-$ & $-$       & $-$ & $-$       & $-$ & $-$ \\
$3S$ & $-0.071$ & $3P$ & $-0.177$ & $3D$ & $0.130$   & $-$ & $-$       & $-$ & $-$       & $-$ & $-$ \\
$4S$ & $-0.109$ & $4P$ & $-0.243$ & $4D$ & $0.183$   & $4F$ & $0.027$  & $-$ & $-$       & $-$ & $-$ \\
$5S$ & $-0.129$ & $5P$ & $-0.276$ & $5D$ & $0.203$   & $5F$ & $0.036$  & $5G$ & $0.009$  & $-$ & $-$ \\
$6S$ & $-0.141$ & $6P$ & $-0.295$ & $6D$ & $0.215$   & $6F$ & $0.044$  & $6G$ & $0.014$  & $6H$ & $0.004$ \\
\hline
\hline
\end{tabular}
\end{minipage}
\end{center}
\end{table}
\end{center}

\section{EVALUATION OF $B_{60}$}
\label{evalB60}

Before we give the general result for $B_{60}$ of 
$P$ states, we would like to recall the known
results for the other coefficients in Eq.~(\ref{defH}).
Specifically, we have for the spin-independent double-logarithmic term,
$B_{62}(nP) = \frac{4}{27} \frac{n^2 - 1}{n^2}$
(see Refs.~\cite{Ka1996,JeNa2002}).
Furthermore, the logarithmic terms~\cite{JeCzPa2005} are
$B_{61}(nP_{1/2}) = \frac43\, N(nP) +
\frac{n^2 - 1}{n^2}
\left(\frac{166}{405} -\frac{8}{27} \, \ln 2 \right)$
and $B_{61}(nP_{3/2}) = \frac43\, N(nP) +
\frac{n^2 - 1}{n^2}
\left(\frac{31}{405} -\frac{8}{27} \, \ln 2 \right)$.
Numerical values for $N(nP)$ can be found in Eq.~(17) of
Ref.~\cite{Je2003jpa}.
Based on Eq.~(8.1) of Ref.~\cite{JeCzPa2005} and on the 
results for $b_L$ obtained in this paper as well as
standard analytic techniques for the evaluation of matrix
elements, we are now in the position to give complete
results for the $B_{60}$ coefficients of $P$ states
(see also Table~\ref{table4}),
\begin{subequations}
\label{b60P}
\begin{align}
\label{b60P12}
& B_{60}(nP_{1/2}) =
-\frac{27517}{25920} - \frac{209}{288\,n} + \frac{1223}{960 \, n^2} 
+ \frac{4}{27}  \frac{n^2 - 1}{n^2} \, \ln^2(2)
- \frac{38}{81} \frac{n^2 - 1}{n^2} \, \ln(2)
\nonumber\\[2ex]
& \quad + \left( \frac{25}{6} + \frac{3}{2 \, n} - \frac{9}{2 \,n^2} \right) \,
\zeta(2) \ln(2)
+ \left( -\frac{9151}{10800} -\frac{1}{4\,n} + \frac{1009}{1200 \, n^2}
\right) \, \zeta(2)
+ \left( -\frac{25}{24} -\frac{3}{8\,n} + \frac{9}{8 \, n^2}
\right) \, \zeta(3)
\nonumber\\[2ex]
& \quad + \beta_4(nP_{1/2}) + \beta_5(nP_{1/2}) + 
\left[ \frac{38}{45} - \frac43\,\ln(2) \right] \, 
N(nP) + b_L(nP)\,,
\\[2ex]
\label{b60P32}
& B_{60}(nP_{3/2}) =
-\frac{73321}{103680} + \frac{185}{1152\,n} + \frac{8111}{25920\,n^2}
+ \frac{4}{27}  \frac{n^2 - 1}{n^2} \, \ln^2(2)
- \frac{11}{81} \frac{n^2 - 1}{n^2} \, \ln(2)
\nonumber\\[2ex]
& \quad 
+ \left( \frac{299}{80} - \frac{3}{8 \, n} - \frac{53}{20 \,n^2} \right) \,
\zeta(2) \ln(2)
+ \left( -\frac{24377}{21600} + \frac{1}{16\,n} - \frac{3187}{3600 \, n^2}
\right) \, \zeta(2)
+ \left( -\frac{299}{320} + \frac{3}{32\,n} + \frac{53}{80 \, n^2}
\right) \, \zeta(3)
\nonumber\\[2ex]
& \quad + \beta_4(nP_{3/2}) + \beta_5(nP_{3/2}) + 
\left[ \frac{38}{45} - \frac43\,\ln(2) \right] \, 
N(nP) + b_L(nP)\,.
\end{align}
\end{subequations}
We see that the total value of $B_{60}$ is the sum of high-energy
operators given by inverse powers of the principal quantum
number, linear combinations of terms proportional to $\ln(2)$
and $\zeta$ functions of various arguments, and low-energy
terms $\beta_4$ and $\beta_5$ which are known from
one-loop calculations~\cite{Je2003jpa,JeEtAl2003}
(see also Tables~\ref{table7} and~\ref{table8} below), as
well as the two-loop Bethe logarithm $\overline{b}_L$.

For $D$ states, it is known that $B_{62}(nD) = B_{61}(nD) = 0$.
Indeed, the fact that $N(nD) = 0$ for $D$ states implies that the 
term of order $1/k_1$ in the asymptotics (\ref{k1asymp}) vanishes,
and this is consistent with the zero result for $B_{61}(nD)$.
The nonlogarithmic terms read [note in particular $N(nD) = 0$],
\begin{subequations}
\label{b60D}
\begin{align}
\label{b60D32}
& B_{60}(nD_{3/2}) =
-\frac{125863}{2016000} - \frac{1021}{9600\,n} + \frac{18811}{60480 \, n^2} 
+ \left( \frac{1387}{8400} + \frac{9}{40 \, n} - \frac{313}{420 \,n^2} \right) \,
\zeta(2) \ln(2)
\\
& + \left( -\frac{5501}{151200} - \frac{3}{80\,n} + \frac{1073}{7560 \, n^2}
\right) \, \zeta(2)
+ \left( -\frac{1387}{33600} -\frac{9}{160\,n} + \frac{313}{1680 \, n^2}
\right) \, \zeta(3)
+ \beta_4(nD_{3/2}) + \beta_5(nD_{3/2}) + b_L(nD)\,,
\nonumber\\[2ex]
\label{b60D52}
& B_{60}(nD_{5/2}) =
\frac{61133}{6804000} + \frac{949}{21600\,n} - \frac{4967}{30240\,n^2}
+ \left( \frac{43}{3150} - \frac{1}{10 \, n} + \frac{61}{210 \,n^2} \right) \,
\zeta(2) \ln(2)
\\
& + \left( -\frac{421}{37800} + \frac{1}{60\,n} - \frac{29}{945 \, n^2}
\right) \, \zeta(2)
+ \left( -\frac{43}{12600} + \frac{1}{40\,n} - \frac{61}{840 \, n^2}
\right) \, \zeta(3)
+ \beta_4(nD_{5/2}) + \beta_5(nD_{5/2}) + b_L(nD)\,.
\nonumber
\end{align}
\end{subequations}
\end{widetext}
For $F$ states, 
the values of $B_{60}$ are displayed
in Table~\ref{table6} (detailed 
formulas are given in Appendix~\ref{miscFGH}).
We observe that the total value of $B_{60}$
for $F$ states is numerically rather close to the 
value of $\delta^2 B_{60}$ for each state
as given in Table~\ref{table2}.
Contributions from the fine-structure dependent terms as well as 
those from the high-energy operators gradually vanish as the
angular momentum increases, and the dominant remaining
contribution then stems from $\delta^2 B_{60}$.

For $G$ and $H$ states, the total value
of $B_{60}$ is the same as the value of 
$\delta^2 B_{60}$ up to the level of $\pm 1.0 \times 10^{-3}$
in units of $B_{60}$, as shown in  
Tables~\ref{table3} and~\ref{table6}. 
We may now use our experience regarding the asymptotic behavior of 
one-loop Bethe logarithms~\cite{JeMo2005bethe} and of 
relativistic Bethe logarithms~\cite{JeEtAl2003},
and based on the data in Tables~\ref{table4},~\ref{table5}
and~Table~\ref{table6} for an extrapolation of $B_{60}$ to an arbitrary 
state. Specifically, we conjecture that for a given 
hydrogenic energy level which reads $nL_j$ in the usual
spectroscopic notation,
\begin{align}
\label{conj}
B_{60}(nL_j) \approx & \; \delta^2 B_{60} \approx
\frac{1}{L^3} \, \left(1.7 - \frac{2.0}{n-L+1} \right) 
\pm 50\,\% \,, \nonumber\\[2ex]
& \mbox{for $L \geq 3$, and both $j = L \pm 1/2$}\,.
\end{align}
This conjecture implies that the 
dominant contribution to $B_{60}$ comes 
from the squared decay rates, i.e.~from the 
term $\delta^2 B_{60}$, and the conjecture (\ref{conj})
can be used to estimate this coefficient
for an arbitrary Rydberg state of high angular
momentum.

\begin{center}
\begin{table}
\begin{center}
\begin{minipage}{8.6cm}
\caption{\label{table4}
Values of the total $B_{60}$ coefficient for $P$
states, derived from Eq.~(\ref{b60P}). 
For the numerical values of $\beta_4$ and $\beta_5$,
see Tables~\ref{table8} and~\ref{table9} in Appendix~\ref{miscP}.}
\begin{tabular}{c@{\hspace{0.3cm}}c@{\hspace{0.7cm}}%
c@{\hspace{0.3cm}}c}
\hline
\hline
\rule[-2mm]{0mm}{6mm}
level & $B_{60}$ & 
level & $B_{60}$ \\
\hline
$2P_{1/2}$ & $-1.6(3)$ & $2P_{3/2}$ & $-1.8(3)$ \\
$3P_{1/2}$ & $-2.0(3)$ & $3P_{3/2}$ & $-2.2(3)$ \\
$4P_{1/2}$ & $-2.4(3)$ & $4P_{3/2}$ & $-2.5(3)$ \\
$5P_{1/2}$ & $-2.4(3)$ & $5P_{3/2}$ & $-2.5(3)$ \\
$6P_{1/2}$ & $-2.5(3)$ & $6P_{3/2}$ & $-2.6(3)$ \\
\hline
\hline
\end{tabular}
\end{minipage}
\end{center}
\end{table}
\end{center}

\begin{center}
\begin{table}
\begin{center}
\begin{minipage}{8.6cm}
\caption{\label{table5}
Same as Table~\ref{table4}, but for $D$ states.
The numerical values of $\beta_4$ and $\beta_5$,
which are needed for the numerical evaluation of the 
expressions in Eq.~(\ref{b60D}),
are recorded in Tables~\ref{table9} and~\ref{tableA} 
in Appendix~\ref{miscD}.}
\begin{tabular}{c@{\hspace{0.3cm}}c@{\hspace{0.7cm}}%
c@{\hspace{0.3cm}}c}
\hline
\hline
\rule[-2mm]{0mm}{6mm}
level & $B_{60}$ &
level & $B_{60}$ \\
\hline
$3D_{3/2}$ & $0.141(2)$ & $3D_{5/2}$ & $0.123(2)$ \\
$4D_{3/2}$ & $0.199(2)$ & $4D_{5/2}$ & $0.178(2)$ \\
$5D_{3/2}$ & $0.219(3)$ & $5D_{5/2}$ & $0.196(3)$ \\
$6D_{3/2}$ & $0.230(4)$ & $6D_{5/2}$ & $0.207(4)$ \\
\hline
\hline
\end{tabular}
\end{minipage}
\end{center}
\end{table}
\end{center}

\begin{center}
\begin{table}
\begin{center}
\begin{minipage}{8.6cm}
\caption{\label{table6}
Same as Tables~\ref{table4} and~\ref{table5}, but for $F$,
$G$ and $H$ states.
For these states, the total values of $B_{60}$ are 
close to the numerical values for 
$\delta^2 B_{60}$ as listed in Table~\ref{table3}, and 
the high-energy operators as well as the fine-structure
dependent corrections do not introduce any significant numerical
deviation of the total value of $B_{60}$ 
from $\delta^2 B_{60}$.}
\begin{tabular}{c@{\hspace{0.3cm}}c@{\hspace{0.7cm}}%
c@{\hspace{0.3cm}}c}
\hline
\hline
\rule[-2mm]{0mm}{6mm}
level & $B_{60}$ &
level & $B_{60}$ \\
\hline
$4F_{5/2}$ & $0.027(1)$ & $4F_{7/2}$ & $0.023(1)$ \\
$5F_{5/2}$ & $0.037(1)$ & $5F_{7/2}$ & $0.033(1)$ \\
$6F_{5/2}$ & $0.046(1)$ & $6F_{7/2}$ & $0.041(1)$ \\
\hline
$5G_{7/2}$ & $0.009(1)$ & $5G_{9/2}$ & $0.008(1)$ \\
$6G_{7/2}$ & $0.014(1)$ & $6G_{9/2}$ & $0.013(1)$ \\
\hline
$6H_{9/2}$ & $0.004(1)$ & $6H_{11/2}$ & $0.004(1)$ \\
\hline
\hline
\end{tabular}
\end{minipage}
\end{center}
\end{table}
\end{center}

%
%
\section{CONCLUSIONS}
\label{conclu}

\begin{figure}[ht]
\begin{center}
\begin{minipage}{8.6cm}
\begin{center}
\includegraphics[width=0.7\linewidth]{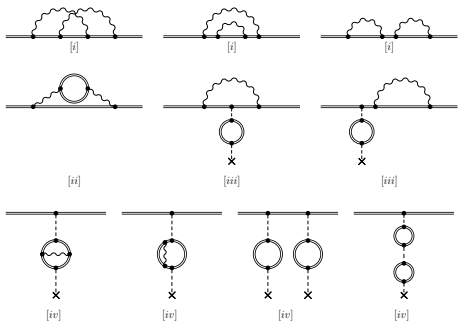}
\caption{\label{fig2} The two-loop diagrams which 
give rise to the $B_{60}$ coefficient naturally 
fall into four gauge-invariant subsets, labeled 
$i$--$iv$ (we follow the classification of diagrams 
according to Ref.~\cite{JeCzPa2005}). 
Our results for $B_{60}$ contain contributions
from all of these diagrams; the separation into the 
four subsets is only relevant for the detailed 
formulas given in the Appendixes.}
\end{center}
\end{minipage}
\end{center}
\end{figure}

Together with the one-loop results from Appendix~\ref{vp},
the reported calculation of the two-loop nonlogarithmic 
term of order $(\alpha/\pi)^2\,(Z \alpha)^2 \, m_{\rm e}\, c^2$ completes
the analysis of the quantum electrodynamic corrections 
to non-$S$ states in the order $\alpha^8 \, m_{\rm e}\, c^2$ for hydrogen
($Z = 1$). Together with other recent theoretical 
investigations~\cite{CzJePa2005,JeCzPa2005}, the current 
work clarifies further prerequisites and the 
theoretical basis for the determination
of fundamental constants from hydrogen spectroscopy at the 
level of one part in $10^{14}$. For ionized helium, the 
corrections calculated here are enhanced by a factor of $2^6 = 64$
in frequency units. The numerical values of principal-value
two-loop Bethe logarithms $\overline{b}_L$
are given in Table~\ref{table2}, and squared decay rates 
$\delta^2 B_{60}$ are indicated in Table~\ref{table3}.
Final numerical results for $B_{60}$, which is the sum 
of $\overline{b}_L$, $\delta^2 B_{60}$ and high-energy 
operators as well as fine-structure dependent terms 
[see Eqs.~(\ref{b60P12})---(\ref{b60D52})], are given in 
Tables~\ref{table4},~\ref{table5} and~\ref{table6}.
Our total results for $B_{60}$ contain the sum of the contributions
from all four gauge-invariant subsets as shown in Fig.~\ref{fig2}.
We confirm that the trend already observed in the 
literature~\cite{DrSw1990,JeMo2005bethe} for 
one-loop Bethe logarithms which decrease in magnitude 
with increasing orbital angular momentum
quantum number, also holds for the two-loop counterparts.
Finally, a comparison of our results with a very recent
numerical investigation~\cite{YeInSh2006} indicates 
that our results for $B_{60}(2P_{1/2})$ and $B_{60}(2P_{3/2})$
for subset $i$ (see Fig.~\ref{fig2} and Tables~\ref{table7} 
and~\ref{table8} below) are consistent with a nonperturbative
(in $Z\alpha$) treatment of the two-loop self-energy, which 
forms subset $i$ in the classification of Fig.~\ref{fig2}.

The connection of the two-loop problem to squared decay rates
has been analyzed previously in Refs.~\cite{JeEvKePa2002,Je2004b60}.
For the atomic levels under investigation, we give a respective 
treatment in Sec.~\ref{square}. The contributions 
to $B_{60}$ in Table~\ref{table3} due to squared imaginary parts
cannot be interpreted as energy shifts in the usual
sense and set a limit to the actual ``definability''
of the atomic energy levels under investigation. 
The surprising conclusion of the current investigation
is that the squared decay rates actually give the dominant
contribution to $B_{60}$ for states with higher angular 
momenta (see Tables~\ref{table2},~\ref{table3} and~\ref{table6}).
As $\delta^2 B_{60}$ cannot be associated uniquely as an
energy correction to one specific atomic level, we conclude 
that our calculation explores the predictive limits of 
quantum electrodynamic theory of energy levels of 
the hydrogen atom and hydrogenlike systems.
Explicit theoretical values of the Lamb shift for selected states in 
hydrogenlike ions will be presented elsewhere.

All calculations reported here (see in particular
Tables~\ref{table2} and~\ref{table3}) 
were carried out on an average-size cluster
of workstations. 
A fifth-order finite-difference scheme was used in order to accurately 
describe the behavior of the $S$ component of the propagators on 
a grid near the origin, and in order to allow for a fast calculation
of the two-loop integrand on the workstation cluster of
Max--Planck--Institute for Nuclear Physics in Heidelberg and 
of the National Institute of Standards and Technology. 
However, even the most sophisticated computer hardware 
and numerical algorithms
would be useless in the current context unless combined with a 
thorough formulation of the subtraction procedures necessary to 
extract the physically observable energy corrections
and frequency shifts.

%
%
\acknowledgments

The author acknowledges support from the 
Deutsche Forschungsgemeinschaft (Heisenberg program).
Helpful conversations with Krzysztof
Pachucki, Vladimir A. Yerokhin and Peter J. Mohr are gratefully acknowledged.
The stimulating atmosphere at the National Institute
of Standards and Technology has contributed to
the completion of this project.

\appendix

\section{MISCELLANEOUS RESULTS FOR $P$ STATES}
\label{miscP}

In Sec.~\ref{evalB60}, we had discussed the evaluation of 
the total $B_{60}$ coefficient for $P_{1/2}$ and $P_{3/2}$ states
of general $n$. Here, we would like to supplement the results
for the individual subsets $i$--$iv$ of the two-loop diagrams 
as given in Figs.~1--4 of Ref.~\cite{JeCzPa2005} and
in Fig.~\ref{fig2} here. The
results presented in this Appendix might be helpful both for a
comparison of our analytic to nonperturbative 
numerical results, as well as for 
a verification of alternative, independent analytic calculations
for particular sets of diagrams.

We start with the subset $i$ (pure two-loop self-energy) for 
$P_{1/2}$ states (see Fig.~\ref{fig2}),
\begin{widetext}
\begin{align}
& B^i_{60}(nP_{1/2}) =
\frac{3829}{77760} + \frac{89}{96\,n} - \frac{44437}{77760 \, n^2}
+ \frac{4}{27}  \frac{n^2 - 1}{n^2} \, \ln^2(2)
- \frac{38}{81} \frac{n^2 - 1}{n^2} \, \ln(2)
\nonumber\\[2ex]
& \quad + \left( \frac{25}{6} + \frac{3}{2 \, n} - \frac{9}{2 \,n^2} \right) \,
\zeta(2) \ln(2)
+ \left( -\frac{7613}{5400} - \frac{5}{4\,n} + \frac{1663}{900 \, n^2}
\right) \, \zeta(2)
+ \left( -\frac{25}{24} - \frac{3}{8\,n} + \frac{9}{8 \, n^2}
\right) \, \zeta(3)
\nonumber\\[2ex]
& \quad + \beta_4(nP_{1/2}) + \beta_5(nP_{1/2}) + 
\left[ \frac{10}{9} - \frac43 \, \ln(2) \right] \, N(nP) +
b_L(nP)\,.
\end{align}

We also note that in Eqs.~(5.4), (6.2) and~(7.7) of Ref.~\cite{JeCzPa2005},
results for the fine-structure difference 
associated with the diagrams of subsets $ii$--$iv$ were indicated.
We reemphasize that these results are valid only for the fine-structure difference
of $P$ states. Therefore, we would like to supplement the individual contributions
for $P_{1/2}$ and $P_{3/2}$ here, starting with $P_{1/2}$,
\begin{align}
B^{ii}_{60}(nP_{1/2}) =&
-\frac{1939}{2160} - \frac{119}{72\,n} + \frac{10577}{6480\,n^2}
+ \left( \frac{9}{16} + \frac{1}{n} - \frac{145}{144\,n^2} \right) \,
\zeta(2) \,,
\\[2ex]
B^{iii}_{60}(nP_{1/2}) =& -\frac{2}{45}\, \left( 1 - \frac{1}{n^2} \right) 
- \frac{4}{15} \, N(nP) \,,
\qquad
B^{iv}_{60}(nP_{1/2}) = -\frac{41}{243}\, \left( 1 - \frac{1}{n^2} \right) \,.
\end{align}

For the two-loop self-energy subset $i$, we obtain in the 
case of a $P_{3/2}$ state,
\begin{align}
& B^i_{60}(nP_{3/2}) =
-\frac{73321}{103680} + \frac{185}{1152\,n} + \frac{8111}{25920\,n^2}
+ \frac{4}{27}  \frac{n^2 - 1}{n^2} \, \ln^2(2)
- \frac{11}{81} \frac{n^2 - 1}{n^2} \, \ln(2)
\nonumber\\[2ex]
& \quad + \left( \frac{299}{80} - \frac{3}{8 \, n} - \frac{53}{20 \,n^2} \right) \,
\zeta(2) \ln(2)
+ \left( -\frac{24377}{21600} + \frac{1}{16\,n} - \frac{3187}{3600 \, n^2}
\right) \, \zeta(2)
+ \left( -\frac{299}{320} + \frac{3}{32\,n} + \frac{53}{80 \, n^2}
\right) \, \zeta(3)
\nonumber\\[2ex]
& \quad + \beta_4(nP_{3/2}) + \beta_5(nP_{3/2}) + 
\left[ \frac{10}{9} - \frac43 \, \ln(2) \right] \, N(nP) +
b_L(nP)\,.
\end{align}
The contributions of the diagrammatic subsets $ii$--$iv$ read as follows,
\begin{align}
B^{ii}_{60}(nP_{3/2}) =&
-\frac{3595}{5184} + \frac{119}{288\,n} - \frac{11}{88\,n^2}
+ \left( \frac{17}{40} - \frac{1}{4\,n} + \frac{7}{90\,n^2} \right) \,
\zeta(2)
\\[2ex]
B^{iii}_{60}(nP_{3/2}) =& -\frac{1}{45}\, \left( 1 - \frac{1}{n^2} \right) 
- \frac{4}{15} \, N(nP) \,,
\qquad
B^{iv}_{60}(nP_{3/2}) = -\frac{41}{486}\, \left( 1 - \frac{1}{n^2} \right)\,.
\end{align}
It is perhaps also instructive to indicate here, the total result for the 
fine-structure difference of $B_{60}$. In Ref.~\cite{JeCzPa2005}, 
only the individual subsets $i$--$iv$ were treated
for the fine-structure, but their sum was not indicated,
and this ``checksum'' is supplemented here:
\begin{align}
& B_{60}(nP_{3/2}) - B_{60}(nP_{1/2})=
\frac{1361}{3840} + \frac{1021}{1152\,n} - \frac{2491}{2592\,n^2}
+ \frac{1}{3} \frac{n^2 - 1}{n^2} \, \ln(2)
+ \left( -\frac{103}{240} - \frac{15}{8 \, n} + \frac{37}{20 \,n^2} \right) \,
\zeta(2) \ln(2)
\\[2ex]
& + \left( -\frac{9}{32} + \frac{5}{16\,n} + \frac{2}{45 \, n^2}
\right) \, \zeta(2)
+ \left( \frac{103}{960} + \frac{15}{32\,n} - \frac{37}{80 \, n^2}
\right) \, \zeta(3)
+ \beta_4(nP_{3/2}) - \beta_4(nP_{1/2}) + \beta_5(nP_{3/2}) - \beta_5(nP_{1/2}) \,.
\nonumber
\end{align}
In this result, $b_L$ cancels, and 
$B_{60}(nP_{3/2}) - B_{60}(nP_{1/2})$ evaluates, e.g., to 
a numerical value of $-0.134817$ for $n=2$. 
The reader may consult Tables~\ref{table7} and~\ref{table8} for details
and observe that the principal numerical uncertainty due to 
$b_L$ cancels for the fine-structure.

\begin{center}
\begin{table}
\begin{center}
\begin{minipage}{14.0cm}
\caption{\label{table7}
Breakdown of the individual contributions to $B_{60}(nP_{1/2})$ from
all diagrammatic subsets as defined in Ref.~\cite{JeCzPa2005}.
The numerical uncertainty in $B^i_{60}(nP_{1/2})$ is entirely due to 
the two-loop Bethe logarithm given in Table~\ref{table2}. 
The $N$ term has been calculated in Ref.~\cite{Je2003jpa};
it is spin-independent and therefore indicated only for
the $P_{1/2}$ states (cf.~Table~\ref{table8} below).}
\begin{tabular}{c@{\hspace{0.3cm}}c@{\hspace{0.3cm}}c@{\hspace{0.3cm}}%
r@{\hspace{0.3cm}}c@{\hspace{0.3cm}}c@{\hspace{0.3cm}}c@{\hspace{0.3cm}}c}
\hline
\hline
\rule[-2mm]{0mm}{6mm}
level & $\beta_4$ & $\beta_5$ & \multicolumn{1}{c}{$N$} &
$B^i_{60}$ & $B^{ii}_{60}$ & $B^{iii}_{60}$ & $B^{iv}_{60}$ \\
\hline
$2P_{1/2}$ & -0.314~100 & 0.260~017 &  0.003~301 & -1.5(3) & 0.017~642 & -0.034~214 & -0.126~543 \\
$3P_{1/2}$ & -0.362~439 & 0.292~936 &  0.003~572 & -1.8(3) & 0.022~297 & -0.040~459 & -0.149~977 \\
$4P_{1/2}$ & -0.380~983 & 0.304~244 & -0.000~394 & -2.2(3) & 0.024~123 & -0.041~562 & -0.158~179 \\
$5P_{1/2}$ & -0.390~543 & 0.309~438 & -0.004~304 & -2.2(3) & 0.025~057 & -0.041~519 & -0.161~975 \\
$6P_{1/2}$ & -0.396~287 & 0.312~249 & -0.007~497 & -2.3(3) & 0.025~613 & -0.041~211 & -0.164~037 \\
\hline
\hline
\end{tabular}
\end{minipage}
\end{center}
\end{table}
\end{center}

\begin{center}
\begin{table}
\begin{center}
\begin{minipage}{14.0cm}
\caption{\label{table8}
Same as Table~\ref{table7}, but for $nP_{3/2}$.}
\begin{tabular}{c@{\hspace{0.3cm}}c@{\hspace{0.3cm}}%
c@{\hspace{0.3cm}}c@{\hspace{0.3cm}}c@{\hspace{0.3cm}}c@{\hspace{0.3cm}}c}
\hline
\hline
\rule[-2mm]{0mm}{6mm}
level & $\beta_4$ & $\beta_5$ &
$B^i_{60}$ & $B^{ii}_{60}$ & $B^{iii}_{60}$ & $B^{iv}_{60}$ \\
\hline
$2P_{3/2}$ & 0.157~050 & -0.130~009 & -1.8(3) & 0.004~207 & 0.015~786 & 0.063~272 \\
$3P_{3/2}$ & 0.181~219 & -0.146~468 & -2.2(3) & 0.005~208 & 0.018~801 & 0.074~989 \\
$4P_{3/2}$ & 0.190~492 & -0.152~122 & -2.6(3) & 0.005~510 & 0.020~938 & 0.079~090 \\
$5P_{3/2}$ & 0.195~271 & -0.154~719 & -2.6(3) & 0.005~627 & 0.022~481 & 0.080~988 \\
$6P_{3/2}$ & 0.198~144 & -0.156~124 & -2.7(3) & 0.005~678 & 0.023~604 & 0.082~019 \\
\hline
\hline
\end{tabular}
\end{minipage}
\end{center}
\end{table}
\end{center}

%
%
\section{Miscellaneous Results for $D$ States}
\label{miscD}

For $D$ states, we observe that only subsets $i$ and $ii$ contribute and 
start with $D_{3/2}$,
\begin{align}
\label{B60iD32}
B^i_{60}(nD_{3/2}) = & \;
\frac{405431}{6048000} + \frac{453}{3200\,n} - \frac{4901}{12096\, n^2} 
+ \left( \frac{1387}{8400} + \frac{9}{40 \, n} - \frac{313}{420 \, n^2} \right) \,
\zeta(2) \ln(2)
\nonumber\\[2ex]
& \; + \left( -\frac{3469}{30240} - \frac{3}{16 \, n} + \frac{4349}{7560 \, n^2} \right) \,
\zeta(2)
+ \left( -\frac{1387}{33600} -\frac{9}{160\,n} + \frac{313}{1680 \, n^2}
\right) \, \zeta(3)
\nonumber\\[2ex]
& \; + \beta_4(nD_{3/2}) + \beta_5(nD_{3/2}) + b_L(nD)\,.
\end{align}
Subset $ii$ yields,
\begin{align}
B^{ii}_{60}(nD_{3/2}) = -\frac{5593}{43200} - \frac{119}{480\,n} + \frac{1547}{2160\,n^2}
+ \left( \frac{47}{600} + \frac{3}{20\,n} - \frac{13}{30\,n^2} \right) \, \zeta(2)\,.
\end{align}
We now continue with $D_{5/2}$,
\begin{align}
\label{B60iD52}
& B^i_{60}(nD_{5/2}) =
-\frac{21503}{756000} - \frac{53}{800\,n} + \frac{1577}{6048\,n^2}
+ \left( \frac{43}{3150} - \frac{1}{10 \, n} + \frac{61}{210 \,n^2} \right) \,
\zeta(2) \ln(2)
\nonumber\\[2ex]
& + \left( \frac{39}{2520} + \frac{1}{12\,n} - \frac{272}{945 \, n^2}
\right) \, \zeta(2)
+ \left( -\frac{43}{12600} + \frac{1}{40\,n} - \frac{61}{840 \, n^2}
\right) \, \zeta(3)
+ \beta_4(nD_{5/2}) + \beta_5(nD_{5/2}) + b_L(nP)\,.
\end{align}
Here, subset $ii$ yields
\begin{align}
B^{ii}_{60}(nD_{5/2}) = \frac{1819}{48600} + 
\frac{119}{1080\,n} - \frac{17}{40\,n^2}
+ \left( -\frac{107}{4725} - 
\frac{1}{15\,n} + \frac{9}{35\,n^2} \right) \, \zeta(2)\,.
\end{align}
Subsets $iii$ and $iv$ do not contribute in either case for $D$ states.
A ``checksum'' for the total fine-structure difference can be useful,
\begin{align}
B_{60}(nD_{5/2}) - B_{60}(nD_{3/2})=& \;
\frac{777473}{10886400} + \frac{2597}{17280\,n} - \frac{5749}{12096\,n^2}
+ \left( -\frac{3817}{25200} - \frac{13}{40 \, n} + \frac{29}{28 \,n^2} \right) \,
\zeta(2) \ln(2)
\nonumber\\[2ex]
& \; + \left( \frac{3817}{151200} + \frac{13}{240\,n} - \frac{29}{168 \, n^2}
\right) \, \zeta(2)
+ \left( \frac{3817}{100800} + \frac{13}{160\,n} - \frac{29}{112 \, n^2}
\right) \, \zeta(3)
\nonumber\\[2ex]
& \; + \beta_4(nD_{5/2}) - \beta_4(nD_{3/2}) + \beta_5(nD_{5/2}) - \beta_5(nD_{3/2}) \,.
\end{align}
In this result, $b_L$ cancels, and the
quantity $B_{60}(nD_{5/2}) - B_{60}(nD_{3/2})$ evaluates, e.g., to
a numerical value of $-0.018197$ for $n=2$.
The interested 
reader may consult Tables~\ref{table9} and~\ref{tableA} for details,
observing that the principal numerical uncertainty due to 
$b_L$ cancels for the fine-structure.

\begin{center}
\begin{table}
\begin{center}
\begin{minipage}{14.0cm}
\caption{\label{table9}
Breakdown of the individual contributions to $B_{60}(nD_{3/2})$ from
all diagrammatic subsets as defined in Ref.~\cite{JeCzPa2005}.
The numerical uncertainty in $B^i_{60}(nD_{3/2})$ is entirely due to
the two-loop Bethe logarithm given in Table~\ref{table2}.}
\begin{tabular}{c@{\hspace{0.3cm}}c@{\hspace{0.3cm}}c@{\hspace{0.3cm}}c@{\hspace{0.3cm}}c}
\hline
\hline
\rule[-2mm]{0mm}{6mm}
level & $\beta_4$ & $\beta_5$ & $B^i_{60}$ & $B^{ii}_{60}$ \\
\hline
$3D_{3/2}$ & -0.002~361 & 0.005~397 & 0.141(2) & -0.000~629 \\
$4D_{3/2}$ & -0.002~883 & 0.006~280 & 0.199(2) & -0.000~696 \\
$5D_{3/2}$ & -0.003~101 & 0.006~675 & 0.220(3) & -0.000~714 \\
$6D_{3/2}$ & -0.003~200 & 0.006~886 & 0.230(4) & -0.000~716 \\
\hline
\hline
\end{tabular}
\end{minipage}
\end{center}
\end{table}
\end{center}

\begin{center}
\begin{table}
\begin{center}
\begin{minipage}{14.0cm}
\caption{\label{tableA}
Same as Table~\ref{table9}, but for $nD_{5/2}$.}
\begin{tabular}{c@{\hspace{0.3cm}}c@{\hspace{0.3cm}}c@{\hspace{0.3cm}}c@{\hspace{0.3cm}}c}
\hline
\hline
\rule[-2mm]{0mm}{6mm}
level & $\beta_4$ & $\beta_5$ & $B^i_{60}$ & $B^{ii}_{60}$ \\
\hline
$3D_{5/2}$ & 0.001~574 & -0.003~598 & 0.123(2) & 0.000~128 \\
$4D_{5/2}$ & 0.001~922 & -0.004~186 & 0.177(2) & 0.000~182 \\
$5D_{5/2}$ & 0.002~067 & -0.004~450 & 0.196(3) & 0.000~202 \\
$6D_{5/2}$ & 0.002~133 & -0.004~591 & 0.207(4) & 0.000~209 \\
\hline
\hline
\end{tabular}
\end{minipage}
\end{center}
\end{table}
\end{center}

%
%
\section{Miscellaneous Results for $F$, $G$ and $H$ States}
\label{miscFGH}

For states with angular momenta $l = 3,4,5$, we indicate here
only the final results for $B_{60}$, without considering the 
breakdown for the terms generated by the individual subsets in 
Fig.~\ref{fig2}. We obtain for $F_{5/2}$ states,
\begin{align}
B_{60}(nF_{5/2}) = & \;
-\frac{8321}{666792} - \frac{1415}{42336\,n} + \frac{2567}{18144\, n^2}
+ \left( \frac{71}{2205} + \frac{1}{14 \, n} - \frac{209}{630 \, n^2} \right) \,
\zeta(2) \ln(2)
\nonumber\\[2ex]
& \; + \left( -\frac{2173}{317520} - \frac{1}{84 \, n} + \frac{347}{5670 \, n^2} \right) \,
\zeta(2)
+ \left( -\frac{71}{8820} - \frac{1}{56\,n} + \frac{209}{2520 \, n^2}
\right) \, \zeta(3)
\nonumber\\[2ex]
& \; + \beta_4(nF_{5/2}) + \beta_5(nF_{5/2}) + b_L(nF)\,.
\end{align}
For $F_{7/2}$ states, the result reads
\begin{align}
B_{60}(nF_{7/2}) = & \;
\frac{994501}{284497920} + \frac{1343}{75264\,n} - \frac{33031}{362880\, n^2}
+ \left( -\frac{125}{56448} - \frac{9}{224 \, n} + \frac{89}{504 \, n^2} \right) \,
\zeta(2) \ln(2)
\nonumber\\[2ex]
& \; + \left( -\frac{5629}{5080320} + \frac{3}{448 \, n} - \frac{1067}{45360 \, n^2} \right) \,
\zeta(2)
+ \left( \frac{125}{225792} + \frac{9}{896\,n} - \frac{89}{2016 \, n^2}
\right) \, \zeta(3)
\nonumber\\[2ex]
& \; + \beta_4(nF_{7/2}) + \beta_5(nF_{7/2}) + b_L(nF)\,.
\end{align}
$G_{7/2}$ states provide us with the following information,
\begin{align}
B_{60}(nG_{7/2}) = & \;
-\frac{762011}{191600640} - \frac{67}{4608\,n} + \frac{2389}{29568\, n^2}
+ \left( \frac{2677}{266112} + \frac{1}{32 \, n} - \frac{1037}{5544 \, n^2} \right) \,
\zeta(2) \ln(2)
\nonumber\\[2ex]
& \; + \left( -\frac{16601}{7983360} - \frac{1}{192\, n} + \frac{3379}{99792 \, n^2} \right) \,
\zeta(2)
+ \left( -\frac{2677}{1064448} - \frac{1}{128\,n} + \frac{1037}{22176\, n^2}
\right) \, \zeta(3)
\nonumber\\[2ex]
& \; + \beta_4(nG_{7/2}) + \beta_5(nG_{7/2}) + b_L(nG)\,,
\end{align}
whereas $G_{9/2}$ states yield
\begin{align}
B_{60}(nG_{9/2}) = & \;
\frac{170587}{112266000} + \frac{193}{21600\,n} - \frac{1277}{22176\, n^2}
+ \left( -\frac{44}{23625} - \frac{1}{50 \, n} + \frac{73}{630 \, n^2} \right) \,
\zeta(2) \ln(2)
\nonumber\\[2ex]
& \; + \left( -\frac{1153}{12474000} + \frac{1}{300 \, n} - \frac{1037}{62370 \, n^2} \right) \,
\zeta(2)
+ \left( \frac{11}{23625} + \frac{1}{200\,n} - \frac{73}{2520 \, n^2}
\right) \, \zeta(3)
\nonumber\\[2ex]
& \; + \beta_4(nG_{9/2}) + \beta_5(nG_{9/2}) + b_L(nG)\,.
\end{align}
The results for $H_{9/2}$
\begin{align}
B_{60}(nH_{9/2}) = & \;
-\frac{9169711}{5606172000} - \frac{2203}{290400\,n} + \frac{19361}{370656 \, n^2}
+ \left( \frac{14411}{3539250} + \frac{9}{550 \, n} - \frac{1543}{12870 \, n^2} \right) \,
\zeta(2) \ln(2)
\nonumber\\[2ex]
& \; + \left( -\frac{104891}{127413000} - \frac{3}{1100\, n} + 
  \frac{1241}{57915 \, n^2} \right) \,
\zeta(2)
+ \left( -\frac{14411}{14157000} - \frac{9}{2200\,n} + \frac{1543}{51480 \, n^2}
\right) \, \zeta(3)
\nonumber\\[2ex]
& \; + \beta_4(nH_{9/2}) + \beta_5(nH_{9/2}) + b_L(nH)\,,
\end{align}
and $H_{11/2}$
\begin{align}
B_{60}(nH_{11/2}) = & \;
\frac{4047937}{5381925120} + \frac{2131}{418176\,n} - \frac{29401}{741312\, n^2}
+ \left( -\frac{85}{75504} - \frac{1}{88 \, n} + \frac{419}{5148 \, n^2} \right) \,
\zeta(2) \ln(2)
\nonumber\\[2ex]
& \; + \left( \frac{877}{20386080} + \frac{1}{528 \, n} - 
  \frac{1123}{92664 \, n^2} \right) \, \zeta(2)
+ \left( \frac{85}{302016} + \frac{1}{352\,n} - \frac{419}{20592\, n^2}
\right) \, \zeta(3)
\nonumber\\[2ex]
& \; + \beta_4(nH_{11/2}) + \beta_5(nH_{11/2}) + b_L(nH)
\end{align}
complete the investigation of $B_{60}$ coefficients.

\end{widetext}

%
%
\section{Vacuum Polarization and Self Energy Shifts
of Order $\boldsymbol{\alpha \, (Z\alpha)^7}$}
\label{vp}

As is well known, 
the one-loop (1L) vacuum polarization (VP) correction scales as
\begin{equation}
\Delta E^{\rm (1L)}_{\mathrm{VP}} = \frac{\alpha}{\pi}\,
\frac{(Z\alpha)^4}{n^3} \, F_{\rm VP}(Z\alpha)\,,
\end{equation}
where $F_{\rm VP}(Z\alpha)$ is a dimensionless function.
The first few terms of $F_{\rm VP}(Z\alpha)$ for $S$ states
can be found in a number of review articles,
e.g.~\cite{SaYe1990,EiGrSh2001}. E.g., the lowest-order term
for $S$ states
is $A^{\mathrm{VP}}_{40}(nS) = -4/15$. The first index
of the $A^{\mathrm{VP}}$-coefficients denotes the power of 
$Z\alpha$, and the second denotes the power of $\ln[(Z\alpha)^{-2}]$.

For states with nonvanishing 
angular momenta, it is well known that the leading terms 
in the expansion of $F_{\rm VP}(Z\alpha)$ vanish.
In the order $\alpha(Z\alpha)^6$, the leading coefficients
for $P$ states~\cite{JeSoMo1997} are known to read as follows,
\begin{subequations}
\label{A60VP}
\begin{align}
A^{\rm VP}_{60}(nP_{1/2}) =& \; - \frac{3}{35}\, \frac{n^2 - 1}{n^2}\,,
\\[2ex]
A^{\rm VP}_{60}(nP_{3/2}) =& \; - \frac{2}{105}\, \frac{n^2 - 1}{n^2}\,.
\end{align}
\end{subequations}
An investigation of the behavior of the wave functions
near the nucleus, inspired by Ref.~\cite{Sc1970},
leads to the following correction terms of order $\alpha(Z\alpha)^7$,
\begin{subequations}
\label{A70VP}
\begin{align}
A^{\rm VP}_{70}(nP_{1/2}) =& \; \frac{41 \pi}{2304}\, \frac{n^2 - 1}{n^2}\,,
\\[2ex]
A^{\rm VP}_{70}(nP_{3/2}) =& \; \frac{7 \pi}{768}\, \frac{n^2 - 1}{n^2}\,.
\end{align}
\end{subequations}
These results are consistent with the particular case $n=2$ treated 
on p.~124 of Ref.~\cite{EiGrSh2001}, for which the coefficients
read $A_{70}(2P_{1/2}) = 41 \pi/3072$ and
$A_{70}(2P_{3/2}) = 7 \pi/1024$. Both the 
$A^{\rm VP}_{60}$ as well as the $A^{\rm VP}_{70}$ coefficients vanish 
for $D$ states and all states with higher angular momenta.
Note that the self-energy remainder function $G_{\rm SE}$, as calculated
in Refs.~\cite{JeMoSo2001pra,JeMo2005p} for $P$ states, 
contains all contributions of order $\alpha \, (Z\alpha)^7$ 
due to the one-photon self-energy. For $D$ states and states with higher
angular momenta, self-energy shifts of order $\alpha \, (Z\alpha)^6$
have been compiled, e.g., 
in Ref.~\cite{Je2005mpla}, and the self-energy corrections 
of order $\alpha \, (Z\alpha)^7$ vanish. These remarks
supplement the above results for the one-loop vacuum polarization.

\end{document}